\documentclass[english,preprint,aps,prc,tightenlines,superscriptaddress]{revtex4}
\usepackage[T1]{fontenc}
\usepackage[latin1]{inputenc}
\usepackage{array}
\usepackage{graphicx}

\makeatletter

\providecommand{\tabularnewline}{\\}

\usepackage{babel}
\makeatother
\begin{document}

\title{Electroweak Hard Photon Bremsstrahlung in Electron-Nucleon Scattering}

\author{A. Aleksejevs}

\affiliation{Department of Astronomy and Physics, Saint Mary's University, Halifax,
NS, Canada}

\author{S. Barkanova}

\affiliation{Department of Physics, Acadia University, Wolfville, NS, Canada}

\author{P.\ G.\ Blunden}

\affiliation{Department of Physics and Astronomy, University of Manitoba, Winnipeg,
MB, Canada R3T 2N2}

\author{N. Deg}

\affiliation{Department of Astronomy and Physics, Saint Mary's University, Halifax,
NS, Canada}

\date{\today}

\begin{abstract}
One way to treat the infrared divergences of the electroweak Next-to-Leading-Order
(NLO) differential cross sections to parity-violating (PV) electron-proton
scattering is by adding soft-photon emission contribution. Although
more physical, the results are left with a logarithmic dependence
on the photon detector acceptance, which can only be eliminated by
considering Hard Photon Bremsstrahlung (HPB) contribution. Here we
present a treatment of HPB for PV electron-proton scattering. HPB
differential cross sections for electron-proton scattering have been
computed using the experimental values of nucleon form factors. The
final results are expressed through kinematic parameters, making it
possible to apply the computed PV HPB differential cross sections
for the analysis of data of a range of current and proposed experiments. 
\end{abstract}
\maketitle

\section{Introduction}

Electroweak properties of the nucleon can be studied by parity-violating
electron-nucleon scattering at low to medium energies \cite{Bec89}.
Such experiments can measure the asymmetry factor coming from the
difference between cross sections of left- and right-handed electrons.
This asymmetry between left- and right-handed particles, as a result
of a parity-violating interference between the weak and electromagnetic
forces, is clearly predicted in the Standard Model of Particle Physics.

Extracting the physics of interest from the measured asymmetry requires
evaluating NLO contribution to electroweak scattering at very high
precision. The method for evaluation of the electron-nucleon up to
NLO differential cross sections most commonly found in the literature
is to follow the Feynman rules for the particles of the Standard Model.
The dominant contribution normally comes from the leading order (LO)
correction in perturbation theory. Some of the electroweak NLO contributions
to intermediate energy, parity non-conserving semi-leptonic neutral
current interactions have been addressed previously in \cite{Mus89,MH90,MH91,Mus94}.
Ref. \cite{Zhu00} also estimated effects due to an intrinsic weak
interaction in the nucleon (e. g. the anapole moment) in chiral perturbation
theory, and found the anapole moment contributions insignificant,
only slightly enhancing the axial vector NLO contribution.

Later work in \cite{BAB2002} took the advantage of the modern computational
opportunities and improved the techniques for one-quark NLO computation
by retaining analytical momentum-dependent expressions, and providing
the numerical evaluations of 446 one-loop diagrams. It also included
calculation of the soft photon emission contribution. However, in
\cite{BAB2002}, even after removing infra-red (IR) divergences through
soft-photon emission corrections, calculated one-quark NLO contribution
show a logarithmic dependence on the detector's photon acceptance
parameter $\Delta E$.

The article presented here demonstrates that elimination of this dependence
can be achieved by adding the Hard-Photon Bremsstrahlung (HPB) differential
cross section. We express general electroweak couplings by inserting
appropriate form factors into vertices and construct HPB differential
cross sections as a function of Mandelstam invariants. For each set
of experimental constraints, integration over the emitted photon phase
space can be performed numerically. Analytical results of this article
can be used for several recent PV experiments \cite{SAMPLE,SAMPLEIII,HAPPEX,G0,A4,Qweak}.

The article provides a detailed description of both hard- and soft-photon
emission treatment of infrared divergences the PV electroweak interference
and pure weak contributions to the total differential cross section.
As an example, we choose to consider electron-proton scattering, as
one of the most relevant cases from a physics perspective. However,
the same technique of treating infrared (IR) divergences can be expanded
to neutron or any other baryon target if same effective structure
of the coupling is used.

\section{Soft-Photon Bremsstrahlung}

If the structure of the nucleon is investigated using weak, neutral
current probe \cite{G0,SAMPLE,HAPPEX}, it is necessary to enhance
the weak contribution in electron-nucleon scattering by exploiting
the parity-violating nature of the weak interactions and constructing
the following quantity (asymmetry): \begin{eqnarray}
 & A=\frac{d\sigma_{R}^{tot}-d\sigma_{L}^{tot}}{d\sigma_{R}^{tot}+d\sigma_{L}^{tot}}\cong\frac{Re\left(M_{LO}^{\gamma}M_{LO+NLO}^{Z}\right)_{R}-Re\left(M_{LO}^{\gamma}M_{LO+NLO}^{Z}\right)_{L}}{\left|M_{LO}^{\gamma}\right|_{R(L)}^{2}} & =\nonumber \\
\label{eq:pv1}\\ & =A_{LO}+\frac{Re\left(M_{LO}^{\gamma}M_{NLO}^{Z}\right)_{R}-Re\left(M_{LO}^{\gamma}M_{NLO}^{Z}\right)_{L}}{\left|M_{LO}^{\gamma}\right|_{R(L)}^{2}}.\nonumber \end{eqnarray}
 Here, $A_{LO}$ is a leading order asymmetry measured in the parts
per million (ppm) and second term of Eq.(\ref{eq:pv1}) is a parity-violating
NLO contribution to the asymmetry.

Finally, if the axial-vector ($C_{1}$) or vector-axial ($C_{2}$)
form factors of the parity violating amplitude are studied \cite{Qweak},
which can be taken from PV Hamiltonian \begin{equation}
H^{PV}=\frac{G_{F}}{\sqrt{2}}\left[C_{1}\left(\overline{u}_{e}\gamma^{\mu}\gamma^{5}u_{e}\right)\left(\overline{u}_{N}\gamma^{\mu}u_{N}\right)+C_{2}\left(\overline{u}_{e}\gamma^{\mu}u_{e}\right)\left(\overline{u}_{N}\gamma^{\mu}\gamma^{5}u_{N}\right)\right],\label{eq:PVhamiltonian}\end{equation}
 with perturbative expansion resulting in \begin{equation}
C_{1,2}=C_{1,2}^{LO}+C_{1,2}^{NLO}+O\left(\alpha^{3}\right).\label{i1.5}\end{equation}

\begin{figure}[h]
\includegraphics[scale=0.5]{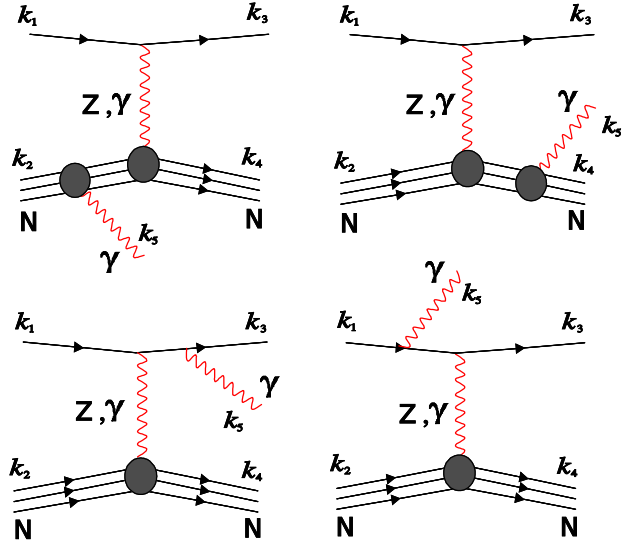} 

\caption{Hard Photon Bremsstrahlung diagrams in electron-proton scattering.\label{fbr1.1} }
\end{figure}

All of the above NLO contributions to the either asymmetry (Eq.(\ref{eq:pv1}))
or PV form-factor (Eq.(\ref{i1.5})) in general can be infra-red divergent
\cite{HooftVelt72}, and can be treated by the soft and hard-photon
emission contribution shown on Fig. (\ref{fbr1.1}).

The differential cross section associated with bremsstrahlung emission
in electron-nucleon scattering can be described by the following formula,
\begin{equation}
d\sigma_{B}\propto\left|M_{B}^{\gamma}+M_{B}^{Z}\right|^{2}=\left|M_{B}^{\gamma}\right|^{2}+2Re\left(M_{B}^{\gamma}M_{B}^{Z}\right)+\left|M_{B}^{Z}\right|^{2}.\label{eq:br1.1}\end{equation}
 The first term of Eq.(\ref{eq:br1.1}) is responsible for the cancellation
of IR divergences if a parity conserving electromagnetic probe is
used. The second term, when used in the asymmetry \begin{equation}
A_{B}\propto\frac{Re\left(M_{B}^{\gamma}M_{B}^{Z}\right)_{R}-Re\left(M_{B}^{\gamma}M_{B}^{Z}\right)_{L}}{\left|M_{LO}^{\gamma}\right|_{R(L)}^{2}},\label{eq:asymbr}\end{equation}
 is responsible for canceling IR divergences in Eq.(\ref{eq:pv1}).
Finally, IR divergences in the NLO form-factors (Eq.(\ref{i1.5}))
are indirectly treated by the third term of Eq.(\ref{eq:br1.1}),
and will be discussed later in this article.

Generally, bremsstrahlung diagrams can be described as $2\rightarrow3$
processes in which integration over emitted photon's phase space should
be performed. If the momentum of the emitted photon is small enough
to be neglected in the numerator algebra, we can present the bremsstrahlung
cross section as a soft photon factor multiplied by the tree level
differential cross section of $2\rightarrow2$ process.

Let us consider an example. The scattering amplitude for the first
diagram of Fig.(\ref{fbr1.1}), for the neutral current reaction,
has the following structure: \begin{eqnarray*}
M_{B}^{Z} & = & \left\langle \overline{u}(m_{e},k_{3})\right|\Gamma_{Z-e}^{\nu}\left|u(m_{e},k_{1})\right\rangle \times\\
\\ &  & \quad\left\langle \overline{u}(m_{N},k_{4})\right|\Gamma_{Z-N}^{\mu}(q)\frac{\not k_{2}-\not k_{5}+m_{N}}{(k_{2}-k_{5})^{2}-m_{N}^{2}}\Gamma_{\gamma-N}^{\alpha}(k_{5})\left|u(m_{N},k_{2})\right\rangle \times\\
\\ &  & \quad\frac{g_{\mu\nu}}{(k_{4}-k_{2}+k_{5})^{2}-m_{Z}^{2}}\varepsilon_{\alpha}^{*}(k_{5}),\end{eqnarray*}
 where the photon polarization vector enters as $\varepsilon_{\alpha}(k_{5})$;
$\Gamma_{Z-e}^{\nu},\,\Gamma_{Z-N}^{\mu}$ and $\Gamma_{\gamma-N}^{\alpha}$
are the couplings of electron with Z boson, nucleon with Z boson,
and photon with nucleon, respectively, defined as \begin{eqnarray}
\Gamma_{Z-e}^{\nu} & = & ie\left[-\frac{1-2s_{W}^{2}}{2c_{W}\, s_{W}}\gamma^{\nu}\varpi_{-}+\frac{s_{W}}{c_{W}}\gamma^{\nu}\varpi_{+}\right],\nonumber \\
\nonumber \\\Gamma_{Z-N}^{\mu}\,(q) & = & ie\left[f_{1}(q)\gamma^{\mu}+\frac{i}{2m_{N}}\sigma^{\mu\rho}q_{\rho}\, f_{2}(q)+g_{1}(q)\gamma^{\mu}\gamma_{5}\right],\label{br1.2}\\
\nonumber \\\Gamma_{\gamma-N}^{\alpha}\,(q) & = & ie\left[F_{1}(q)\gamma^{\alpha}+\frac{i}{2m_{N}}\sigma^{\alpha\rho}(q)_{\rho}F_{2}(q)\right].\nonumber \end{eqnarray}
 Here $q$ corresponds to the momentum transferred to the nucleon
from the vector boson. The shortened notation of $s_{W}$ and $c_{W}$
refers to $\sin\theta_{W}$ and $\cos\theta_{W}$, the Weinberg mixing
angle. For the form factors $f_{1,2}(q),$ $F_{1,2}(q),$ and $g_{1}(q)$
we have used \begin{eqnarray}
f_{1,2}(q) & = & \frac{1}{4s_{W}\, c_{W}}\left(F_{1,2}^{V\left(N\right)}-4s_{W}^{2}F_{1,2}\right)G_{N}\left(q\right),\nonumber \\
\nonumber \\F_{1,2}\left(q\right) & = & F_{1,2}(0)G_{N}\left(q\right),\nonumber \\
\label{br1.2a}\\F_{1,2}^{V\left(n\right)}\left(0\right) & = & F_{1,2}^{n}\left(0\right)-F_{1,2}^{p}\left(0\right),\,\,\,\, F_{1,2}^{V\left(p\right)}\left(0\right)=F_{1,2}^{p}\left(0\right)-F_{1,2}^{n}\left(0\right),\nonumber \\
\nonumber \\g_{1}(q) & = & -\frac{g_{A}(0)}{4s_{W}\, c_{W}}G_{N}\left(q\right),\nonumber \end{eqnarray}
 with $F_{1,2}\left(0\right)$ and $g_{A}\left(0\right)$ defined
as the nucleon's Dirac, Pauli, and axial form factors at zero momentum
transfer and corresponds to the electric charge, anomalous magnetic
moment, and axial charge, respectively. Here, we use the universal
formfactor of monopole or dipole type $G_{N}\left(q\right)=\left(\frac{\Lambda^{2}}{\Lambda^{2}-q^{2}}\right)^{n}$
with $n=1,2$ and $\Lambda^{2}=0.83m_{N}^{2}.$

In the soft-photon emission limit the coupling between the emitted
photon and the nucleon, $\Gamma_{\gamma-N}^{\alpha}(k_{5})$, is just
equal to $ieQ\gamma^{\alpha}$, where charge $Q=1(0)$ for proton
(neutron). The numerator of nucleon's propagator $\not k_{2}-\not k_{5}+m_{N}$
can be replaced by $\not k_{2}+m_{N},$ and $(k_{2}-k_{5})^{2}-m_{N}^{2}$
can be easily simplified into $-2\left(k_{2}\cdot k_{5}\right)$.
Using the Dirac equation for free spinors, we have \begin{eqnarray}
\frac{\not k_{2}+m_{N}}{-2\left(k_{2}\cdot k_{5}\right)}ieQ\gamma^{\alpha}\left|u(m_{N},k_{2})\right\rangle \varepsilon_{\alpha}(k_{5}) & = & -ieQ\frac{\not k_{2}\gamma^{\alpha}+\gamma^{\alpha}\not k_{2}}{2\left(k_{2}\cdot k_{5}\right)}\left|u(m_{N},k_{2})\right\rangle \varepsilon_{\alpha}^{*}(k_{5})\nonumber \\
\label{br1.3}\\\  & = & -ieQ\frac{(k_{2}\cdot\varepsilon^{*}(k_{5}))}{(k_{2}\cdot k_{5})}\left|u(m_{N},k_{2})\right\rangle .\nonumber \end{eqnarray}
 Now we can present the soft-photon amplitude in the following form:
\begin{eqnarray}
M_{B}^{soft} & = & \left\langle \overline{u}(m_{e},k_{3})\right|\Gamma_{Z-e}^{\nu}\left|u(m_{e},k_{1})\right\rangle \left\langle \overline{u}(m_{N},k_{4})\right|\Gamma_{Z-N}^{\mu}(q)\left|u(m_{N},k_{2})\right\rangle \times\nonumber \\
\label{br1.4a}\\ &  & \frac{g_{\mu\nu}}{(k_{4}-k_{2}+k_{5})^{2}-m_{Z}^{2}}\left(-ieQ\frac{(k_{2}\cdot\varepsilon^{*}(k_{5}))}{(k_{2}\cdot k_{5})}\right)\nonumber \\
\nonumber \\ & = & M_{LO}\left(-ieQ\frac{(k_{2}\cdot\varepsilon^{*}(k_{5}))}{(k_{2}\cdot k_{5})}\right)=M_{LO}\kappa(k_{2},k_{5}).\nonumber \end{eqnarray}
 Here, $M_{LO}$ is a tree level amplitude of $2\rightarrow2$ process.
Eq.(\ref{br1.4a}) can be used for the other photon emission diagrams
with a different factor $\kappa(k_{i},k_{5})=\pm ieQ_{i}\frac{(k_{i}\cdot\varepsilon^{*}(k_{5}))}{(k_{i}\cdot k_{5})}$.
Now we can sum over all four graphs on Fig.(\ref{fbr1.1}) and square
the total amplitude to get the following: \begin{equation}
\left|M_{B}^{soft}\right|^{2}=\left|M_{LO}\right|^{2}e^{2}\left|\frac{(k_{1}\cdot\varepsilon^{*}(k_{5}))}{(k_{1}\cdot k_{5})}-Q\frac{(k_{2}\cdot\varepsilon^{*}(k_{5}))}{(k_{2}\cdot k_{5})}-\frac{(k_{3}\cdot\varepsilon^{*}(k_{5}))}{(k_{3}\cdot k_{5})}+Q\frac{(k_{4}\cdot\varepsilon^{*}(k_{5}))}{(k_{4}\cdot k_{5})}\right|^{2}.\label{br1.5}\end{equation}
 The photon couples to a current which is conserved: $k^{\mu}M_{\mu}=0.$
This fact, and the summation over all photon polarizations gives us
the possibility to replace $\sum_{\varepsilon}(k_{i})^{\mu}(k_{j})^{\nu}\varepsilon_{\mu}^{*}\varepsilon_{\nu}$
with $-g_{\mu\nu}(k_{i})^{\mu}(k_{j})^{\nu}=-(k_{i}\cdot k_{j})$.
The last step is to integrate over the emitted photon phase space
$d\Gamma_{k_{5}}=\frac{d^{3}k_{5}}{\left(2\pi^{3}\right)\,2k_{5}^{0}}$
and regularize the infrared divergence by assigning to the photon
small rest mass $\lambda.$ This dependence on the rest mass of the
photon will be canceled when added to IR divergent NLO contribution
(see Eqs.(\ref{eq:pv1}, \ref{eq:asymbr})). The resulting soft-photon
emission differential cross section is expressed as \begin{eqnarray}
d\sigma_{soft} & = & d\sigma_{LO}^{2\rightarrow2}\left(-\frac{\alpha^{2}}{2\pi^{2}}\right)\int_{\lambda\leq\left|k_{5}^{0}\right|\leq\Delta E}\frac{d^{3}k_{5}}{2\sqrt{k_{5}^{2}+\lambda^{2}}}\left(\begin{array}{c}
\frac{k_{1}}{(k_{1}\cdot k_{5})}-Q\frac{k_{2}}{(k_{2}\cdot k_{5})}-\\
\\\frac{k_{3}}{(k_{3}\cdot k_{5})}+Q\frac{k_{4}}{(k_{4}\cdot k_{5})}\end{array}\right)^{2}\nonumber \\
\nonumber \\\  & = & d\sigma_{LO}^{2\rightarrow2}\,\left(-\frac{\alpha^{2}}{2\pi^{2}}\right)(2m_{e}^{2}\, I\left(k_{1},k_{1}\right)-(2m_{e}^{2}-t)\, I\left(k_{1},k_{3}\right)+2Q^{2}m_{N}^{2}\, I\left(k_{2},k_{2}\right)-\nonumber \\
\nonumber \\ &  & \ \ Q^{2}(2m_{N}^{2}-t)\, I\left(k_{2},k_{4}\right)-2Q\,(u-m_{e}^{2}-m_{N}^{2})\, I\left(k_{1},k_{4}\right)-2(s-m_{e}^{2}-m_{N}^{2})\, I\left(k_{1},k_{2}\right))\nonumber \\
\nonumber \\\  & = & d\sigma_{LO}^{2\rightarrow2}\cdot\delta_{soft}.\label{br1.6}\end{eqnarray}
 Here $\Delta E$ is the maximum possible energy of emitted photon
for which the soft-photon approximation is still valid. Numerical
analysis leads to a typical constraint $\lambda<\Delta E\leq10^{-3}E_{cms}$.

In Eq.(\ref{br1.6}), $I\left(k_{i},k_{j}\right)=\int_{\left|k_{5}\right|\leq\Delta E}\frac{d^{3}k_{5}}{2\sqrt{k_{5}^{2}+\lambda^{2}}}\frac{1}{(k_{i}\cdot k_{5})(k_{j}\cdot k_{5})}$
is the soft-photon emission integral evaluated earlier by \cite{HooftVelt72},
and equal to \begin{equation}
I\left(k_{i},k_{j}\right)=\frac{2\pi\alpha_{ij}}{\alpha_{ij}^{2}m_{i}^{2}-m_{j}^{2}}\left[\begin{array}{c}
\frac{1}{2}\ln\left(\frac{\alpha_{ij}^{2}m_{i}^{2}}{m_{j}^{2}}\right)\ln\left(\frac{4\Delta E^{2}}{\lambda^{2}}\right)+\frac{1}{4}\ln^{2}\left(\frac{E_{i}-\left|k\right|_{i}}{E_{i}+\left|k\right|_{i}}\right)-\frac{1}{4}\ln^{2}\left(\frac{E_{j}-\left|k\right|_{j}}{E_{j}+\left|k\right|_{j}}\right)+\\
\\Li_{2}\left(1-\frac{\alpha_{ij}}{v_{ij}}\left(E_{i}+\left|k\right|_{i}\right)\right)+Li_{2}\left(1-\frac{\alpha_{ij}}{v_{ij}}\left(E_{i}-\left|k\right|_{i}\right)\right)-\\
\\Li_{2}\left(1-\frac{1}{v_{ij}}\left(E_{j}+\left|k\right|_{j}\right)\right)-Li_{2}\left(1-\frac{1}{v_{ij}}\left(E_{j}-\left|k\right|_{j}\right)\right)\end{array}\right],\label{br1.7}\end{equation}
 where $v_{ij}=\frac{\alpha_{ij}^{2}m_{i}^{2}-m_{j}^{2}}{2\alpha_{ij}E_{i}-E_{j}},$
and $E_{i}$, $\left|k\right|_{j}$ are the fermion's energy and spatial
momentum in center of mass reference frame, correspondingly. The parameter
$\alpha_{ij}$ can be extracted from Table (\ref{SoftTable}).
\begin{table}
\begin{centering}
\begin{tabular}{|p{1.5cm}p{1.5cm}p{1.5cm}p{1.5cm}p{7cm}|}
\hline 
\multicolumn{1}{|p{1.5cm}|}{$i$}&
\multicolumn{1}{p{1.5cm}|}{$j$}&
\multicolumn{1}{p{1.5cm}|}{$m_{i}$}&
\multicolumn{1}{p{1.5cm}|}{$m_{j}$}&
$\alpha_{ij}$ \tabularnewline
\hline
\hline 
\multicolumn{1}{|p{1.5cm}|}{1}&
\multicolumn{1}{p{1.5cm}|}{1}&
\multicolumn{1}{p{1.5cm}|}{$m_{e}$}&
\multicolumn{1}{p{1.5cm}|}{$m_{e}$}&
1 \tabularnewline
\hline 
\multicolumn{1}{|p{1.5cm}|}{2}&
\multicolumn{1}{p{1.5cm}|}{2}&
\multicolumn{1}{p{1.5cm}|}{$m_{N}$}&
\multicolumn{1}{p{1.5cm}|}{$m_{N}$}&
1 \tabularnewline
\hline 
\multicolumn{1}{|p{1.5cm}|}{1}&
\multicolumn{1}{p{1.5cm}|}{3}&
\multicolumn{1}{p{1.5cm}|}{$m_{e}$}&
\multicolumn{1}{p{1.5cm}|}{$m_{e}$}&
$1-\frac{t}{2m_{e}^{2}}+\frac{\sqrt{t^{2}-4tm_{e}^{2}}}{2m_{e}^{2}}$ \tabularnewline
\hline 
\multicolumn{1}{|p{1.5cm}|}{2}&
\multicolumn{1}{p{1.5cm}|}{4}&
\multicolumn{1}{p{1.5cm}|}{$m_{N}$}&
\multicolumn{1}{p{1.5cm}|}{$m_{N}$}&
$1-\frac{t}{2m_{N}^{2}}+\frac{\sqrt{t^{2}-4tm_{N}^{2}}}{2m_{N}^{2}}$ \tabularnewline
\hline 
\multicolumn{1}{|p{1.5cm}|}{1}&
\multicolumn{1}{p{1.5cm}|}{4}&
\multicolumn{1}{p{1.5cm}|}{$m_{e}$}&
\multicolumn{1}{p{1.5cm}|}{$m_{N}$}&
$\frac{m_{e}^{2}+m_{N}^{2}-u+\sqrt{\left(u-m_{e}^{2}-m_{N}^{2}\right)^{2}-4m_{e}^{2}m_{N}^{2}}}{2m_{e}^{2}}$ \tabularnewline
\hline 
\multicolumn{1}{|p{1.5cm}|}{1}&
\multicolumn{1}{p{1.5cm}|}{2}&
\multicolumn{1}{p{1.5cm}|}{$m_{e}$}&
\multicolumn{1}{p{1.5cm}|}{$m_{N}$}&
$\frac{s-m_{e}^{2}-m_{N}^{2}+\sqrt{\left(m_{e}^{2}+m_{N}^{2}-s\right)^{2}-4m_{e}^{2}m_{N}^{2}}}{2m_{e}^{2}}$ \tabularnewline
\hline
\end{tabular}
\par\end{centering}

\caption{Soft photon emission integral parameters of Eq. (2.10) \label{SoftTable}}
\end{table}

The dependence on $\Delta E$ is canceled as a result of adding soft
and hard-photon emission differential cross sections. It is worthwhile
to mention here that $d\sigma_{soft}$ is proportional to $d\sigma_{LO}^{2\rightarrow2}$,
which is either determined by the $Re\left(M_{LO}^{\gamma}M_{LO}^{Z}\right)$
or $\left|M_{LO}^{Z}\right|^{2}$, and hence gives us for $2Re\left(M_{B}^{\gamma}M_{B}^{Z}\right)_{soft}=2Re\left(M_{LO}^{\gamma}M_{LO}^{Z}\right)\cdot\delta_{soft}$
or $\left|M_{B}^{Z}\right|_{soft}^{2}=\left|M_{LO}^{Z}\right|^{2}\cdot\delta_{soft}$.

\section{Hard-Photon Bremsstrahlung}

This section gives details on the evaluation of hard-photon bremsstrahlung
differential cross section. The results are expressed in a form convenient
for further analysis.

\subsection{Electron-Nucleon Scattering}

In the case where the energy of the emitted photon $\left(k_{5}^{0}>\Delta E\right)$
can no longer be neglected in the numerator algebra, we have to account
for all the differences between hard- and soft-photon emission. Besides
the fact that the hard-photon amplitude will have $k_{5}$ in the
numerator, calculations for differential cross section will have to
include helicity matrix elements with extended set of Mandelstam variables.
These matrix elements come from the use of the momentum conservation
law for $2\rightarrow3$ process. Thus, the helicity matrix elements
will depend on the extended set of Mandelstam variables: \begin{eqnarray}
s & = & \left(k_{1}+k_{2}\right)^{2},\,\, s^{\prime}=\left(k_{3}+k_{4}\right)^{2},\nonumber \\
t & = & \left(k_{1}-k_{3}\right)^{2},\,\, t^{\prime}=\left(k_{2}-k_{4}\right)^{2},\label{br1.9}\\
u & = & \left(k_{1}-k_{4}\right)^{2},\,\, u^{\prime}=\left(k_{2}-k_{3}\right)^{2}.\nonumber \end{eqnarray}
 Let us start with the total amplitude for the set of the four graphs
in Fig.(\ref{fbr1.1}): \begin{eqnarray}
 &  & M_{tot,\{ Z,\gamma\}}^{2\rightarrow3}=\left(\begin{array}{c}
\left\langle \overline{u}_{e}(k_{3})\right|\Gamma_{\{ Z,\gamma\}-e}^{\nu}\left|u_{e}(k_{1})\right\rangle \times\\
\\\left\langle \overline{u}_{N}(k_{4})\right|\Gamma_{\{ Z,\gamma\}-N}^{\mu}(t)\frac{\not k_{2}-\not k_{5}+m_{N}}{(k_{2}-k_{5})^{2}-m_{N}^{2}}\Gamma_{\gamma-N}^{\alpha}(k_{5})\left|u_{N}(k_{2})\right\rangle \end{array}\right)\frac{g_{\mu\nu}}{t-m_{\{ Z,\gamma\}}^{2}}\varepsilon_{\alpha}^{*}(k_{5})\nonumber \\
\nonumber \\ &  & +\left(\begin{array}{c}
\left\langle \overline{u}_{e}(k_{3})\right|\Gamma_{\{ Z,\gamma\}-e}^{\nu}\left|u_{e}(k_{1})\right\rangle \times\\
\\\left\langle \overline{u}_{N}(k_{4})\right|\Gamma_{\gamma-N}^{\alpha}(k_{5})\frac{\not k_{4}+\not k_{5}+m_{N}}{(k_{4}+k_{5})^{2}-m_{N}^{2}}\Gamma_{\{ Z,\gamma\}-N}^{\mu}(t)\left|u_{N}(k_{2})\right\rangle \end{array}\right)\frac{g_{\mu\nu}}{t-m_{\{ Z,\gamma\}}^{2}}\varepsilon_{\alpha}^{*}(k_{5})\nonumber \\
\nonumber \\ &  & +\left(\begin{array}{c}
\left\langle \overline{u}_{e}(k_{3})\right|\Gamma_{\gamma-e}^{\alpha}\frac{\not k_{3}+\not k_{5}+m_{e}}{(k_{3}+k_{5})^{2}-m_{e}^{2}}\Gamma_{\{ Z,\gamma\}-e}^{\nu}\left|u_{e}(k_{1})\right\rangle \times\\
\\\left\langle \overline{u}_{N}(k_{4})\right|\Gamma_{\{ Z,\gamma\}-N}^{\mu}(t^{\prime})\left|u_{N}(k_{2})\right\rangle \end{array}\right)\frac{g_{\mu\nu}}{t^{\prime}-m_{\{ Z,\gamma\}}^{2}}\varepsilon_{\alpha}^{*}(k_{5})\label{br1.10}\\
\nonumber \\ &  & +\left(\begin{array}{c}
\left\langle \overline{u}_{e}(k_{3})\right|\Gamma_{\{ Z,\gamma\}-e}^{\nu}\frac{\not k_{1}-\not k_{5}+m_{e}}{(k_{1}-k_{5})^{2}-m_{e}^{2}}\Gamma_{\gamma-e}^{\alpha}\left|u_{e}(k_{1})\right\rangle \cdot\\
\\\cdot\left\langle \overline{u}_{N}(k_{4})\right|\Gamma_{\{ Z,\gamma\}-N}^{\mu}(t^{\prime})\left|u_{N}(k_{2})\right\rangle \end{array}\right)\frac{g_{\mu\nu}}{t^{\prime}-m_{\{ Z,\gamma\}}^{2}}\varepsilon_{\alpha}^{*}(k_{5}).\nonumber \end{eqnarray}
 The evaluation of the interference term $Re(M_{tot,\,\gamma}^{2\rightarrow3}M_{tot,\, Z}^{2\rightarrow3})$
and $\left|M_{tot,\, Z}^{2\rightarrow3}\right|^{2}$is somewhat cumbersome
because it includes calculations of 3136 helicity matrix elements.
Here, $t^{\prime}-m_{Z}^{2}$ can be replaced by $t-m_{Z}^{2}$ due
to the fact that $\left\{ t,t^{\prime}\right\} \ll m_{Z}^{2}$. The
evaluation of the HPB contribution can be further simplified by spliting
the amplitude into two parts: \begin{equation}
M_{tot,\{ Z,\gamma\}}^{2\rightarrow3}=M_{a,\{ Z,\gamma\}}^{2\rightarrow3}+M_{b,\{ Z,\gamma\}}^{2\rightarrow3}.\label{eq:split}\end{equation}
 Here, $M_{a,\{ Z,\gamma\}}^{2\rightarrow3}$ is the total amplitude
with the momentum of the emitted photon $k_{5}$ removed from the
numerator of Eq.(\ref{br1.10}), and $M_{b,\{ Z,\gamma\}}^{2\rightarrow3}$
is everything that is left up to order $O\left(k_{5}\right)$. Now,
the squared amplitude for the neutral current reaction has a simple
form: \begin{equation}
\left|M_{tot,Z}^{2\rightarrow3}\right|^{2}=\left|M_{a,Z}^{2\rightarrow3}\right|^{2}+2Re(M_{a,Z}^{2\rightarrow3}M_{b,Z}^{2\rightarrow3})+\left|M_{b,Z}^{2\rightarrow3}\right|^{2}.\label{br1.11}\end{equation}

The interference term can be calculated as \begin{equation}
Re(M_{tot,\,\gamma}^{2\rightarrow3}M_{tot,\, Z}^{2\rightarrow3})=Re(M_{a,\gamma}^{2\rightarrow3}M_{a,Z}^{2\rightarrow3})+Re(M_{a,\gamma}^{2\rightarrow3}M_{b,Z}^{2\rightarrow3})+Re(M_{b,\gamma}^{2\rightarrow3}M_{a,Z}^{2\rightarrow3})+Re(M_{b,\gamma}^{2\rightarrow3}M_{b,Z}^{2\rightarrow3}).\label{br1.11int}\end{equation}

Applying Dirac equation in $M_{a,\{ Z,\gamma\}}^{2\rightarrow3}$
amplitude, we can simplify our calculations considerably. The first
terms of Eq. (\ref{br1.11}) and (\ref{br1.11int}) can be obtained
from \begin{eqnarray}
M_{a,\{ Z,\gamma\}}^{2\rightarrow3} & = & M_{a,\{ Z,\gamma\}}^{\prime}\left(\frac{G_{N}\left(t^{\prime}\right)\left(k_{1}\cdot\varepsilon^{*}(k_{5})\right)}{\left(k_{1}\cdot k_{5}\right)\left(t^{\prime}-m_{\{ Z,\gamma\}}^{2}\right)}-Q\frac{G_{N}\left(t\right)\left(k_{2}\cdot\varepsilon^{*}(k_{5})\right)}{\left(k_{2}\cdot k_{5}\right)\left(t-m_{\{ Z,\gamma\}}^{2}\right)}-\right.\nonumber \\
\label{br1.12}\\ &  & \left.\frac{G_{N}\left(t^{\prime}\right)\left(k_{3}\cdot\varepsilon^{*}(k_{5})\right)}{\left(k_{3}\cdot k_{5}\right)\left(t^{\prime}-m_{\{ Z,\gamma\}}^{2}\right)}+Q\frac{G_{N}\left(t\right)\left(k_{4}\cdot\varepsilon^{*}(k_{5})\right)}{\left(k_{4}\cdot k_{5}\right)\left(t-m_{\{ Z,\gamma\}}^{2}\right)}\right),\nonumber \end{eqnarray}
 where \begin{equation}
M_{a,\{ Z,\gamma\}}^{\prime}=ie\left\langle \overline{u}(m_{e},k_{3})\right|\Gamma_{\{ Z,\gamma\}-e}^{\nu}\left|u(m_{e},k_{1})\right\rangle \left\langle \overline{u}(m_{N},k_{4})\right|\left(\Gamma_{\{ Z,\gamma\}-N}^{\mu}\right)^{\prime}\left|u(m_{N},k_{2})\right\rangle g_{\mu\nu}.\label{br1.13}\end{equation}
 The term $\left(\Gamma_{\{ Z,\gamma\}-N}^{\mu}\right)^{\prime}$
represents a coupling which was modified in a way so it would no longer
have dependency on the hadron formfactor $G_{N}\left(t\right)$, and
no longer contain momentum of the photon in its Pauli part of coupling:
\begin{eqnarray}
\left(\Gamma_{Z-N}^{\mu}\right)^{\prime} & = & ie\left[f_{1}(0)\gamma^{\mu}+\frac{i}{2m_{N}}\sigma^{\mu\rho}\left(k_{4}-k_{2}\right)_{\rho}\, f_{2}(0)+g_{1}(0)\gamma^{\mu}\gamma_{5}\right],\label{br1.14}\\
\left(\Gamma_{\gamma-N}^{\mu}\right)^{\prime} & = & ie\left[F_{1}(0)\gamma^{\mu}+\frac{i}{2m_{N}}\sigma^{\mu\rho}\left(k_{4}-k_{2}\right)_{\rho}\, F_{2}(0)\right].\label{eq:br1.14gam}\end{eqnarray}
 In Eq.(\ref{br1.13}) and Eq.(\ref{br1.12}), the coupling $\Gamma_{\gamma-\left\{ e,N\right\} }^{\alpha}$
again was replaced by $\left(ieQ\right)\gamma^{\alpha}$. It is straightforward
to see what after the integration over the phase space of the emitted
photon only term $\left|M_{a,Z}^{2\rightarrow3}\right|^{2}$ and $Re(M_{a,\gamma}^{2\rightarrow3}M_{a,Z}^{2\rightarrow3})$
will have a logarithmic dependence on the photon detector acceptance
parameter $\Delta E$. Therefore $\left|M_{a,Z}^{2\rightarrow3}\right|^{2}$and
$Re(M_{a,\gamma}^{2\rightarrow3}M_{a,Z}^{2\rightarrow3})$, when combined
with the soft-photon bremsstrahlung differential cross section, will
be responsible for the cancellation of $Log(4\Delta E^{2})$ term
in Eqs.(\ref{br1.6}) and (\ref{br1.7}).

Further numerical analysis shows that, when integrated, the second
and third terms of Eq.(\ref{br1.11}) and Eq.(\ref{br1.11int}) have
no logarithmic dependence on $\Delta E.$ They both are small compared
to the first term when energy of incident electrons is in the domain
of the current or proposed PV experiments. This simplifies calculations
of PV HPB contribution considerably, since the only first term of
the Eq.(\ref{br1.11}) and Eq.(\ref{br1.12}) has to be considered
in the calculations. Here we provide details on how to calculate first
terms of Eq.(\ref{br1.11}) and (\ref{br1.11int}) explicitly. Although
details of calculations for the rest of the terms are not shown in
this article we have them included in our numerical analysis.

We can write the term $\left|M_{a,Z}^{2\rightarrow3}\right|_{L,R}^{2}$
of Eq.(\ref{br1.11}) in the following form:

\begin{eqnarray}
\left|M_{a,Z}^{2\rightarrow3}\right|_{L,R}^{2} & = & -\left|M_{a,Z}^{\prime}\right|_{L,R}^{2}\cdot\delta_{HPB'}^{Z},\nonumber \\
\nonumber \\\delta_{HPB'}^{Z} & = & \frac{1}{\left(t-m_{Z}^{2}\right)^{2}}\left(G_{N}\left(t^{\prime}\right)^{2}\left(m_{e}^{2}\left(\frac{1}{\left(k_{1}\cdot k_{5}\right)^{2}}+\frac{1}{\left(k_{3}\cdot k_{5}\right)^{2}}\right)-\frac{2m_{e}^{2}-t}{\left(k_{1}\cdot k_{5}\right)\left(k_{3}\cdot k_{5}\right)}\right)+\right.\nonumber \\
\label{br1.15}\\ &  & Q^{2}G_{N}\left(t\right)^{2}\left(m_{N}^{2}\left(\frac{1}{\left(k_{2}\cdot k_{5}\right)^{2}}+\frac{1}{\left(k_{4}\cdot k_{5}\right)^{2}}\right)-\frac{2m_{N}^{2}-t^{\prime}}{\left(k_{2}\cdot k_{5}\right)\left(k_{4}\cdot k_{5}\right)}\right)+\nonumber \\
\nonumber \\ &  & Q\, G_{N}\left(t\right)G_{N}\left(t^{\prime}\right)\left(\frac{m_{e}^{2}+m_{N}^{2}-u}{\left(k_{1}\cdot k_{5}\right)\left(k_{4}\cdot k_{5}\right)}-\frac{s-m_{e}^{2}-m_{N}^{2}}{\left(k_{1}\cdot k_{5}\right)\left(k_{2}\cdot k_{5}\right)}-\right.\nonumber \\
\nonumber \\ &  & \qquad\left.\left.\frac{s^{\prime}-m_{e}^{2}-m_{N}^{2}}{\left(k_{3}\cdot k_{5}\right)\left(k_{4}\cdot k_{5}\right)}+\frac{m_{e}^{2}+m_{N}^{2}-u^{\prime}}{\left(k_{2}\cdot k_{5}\right)\left(k_{3}\cdot k_{5}\right)}\right)\right).\nonumber \end{eqnarray}
 As for the interference term $Re(M_{a,\gamma}^{2\rightarrow3}M_{a,Z}^{2\rightarrow3})_{L,R}$,
we have: \begin{eqnarray}
Re(M_{a,\gamma}^{2\rightarrow3}M_{a,Z}^{2\rightarrow3})_{L,R} & = & -Re(M_{a,\gamma}^{\prime}M_{a,Z}^{\prime})_{L,R}\cdot\delta_{HPB'}^{Z-\gamma},\nonumber \\
\nonumber \\\delta_{HPB'}^{Z-\gamma} & = & \frac{1}{t-m_{Z}^{2}}\left(\frac{G_{N}\left(t^{\prime}\right)^{2}}{t^{\prime}}\left(m_{e}^{2}\left(\frac{1}{\left(k_{1}\cdot k_{5}\right)^{2}}+\frac{1}{\left(k_{3}\cdot k_{5}\right)^{2}}\right)-\frac{2m_{e}^{2}-t}{\left(k_{1}\cdot k_{5}\right)\left(k_{3}\cdot k_{5}\right)}\right)+\right.\nonumber \\
\nonumber \\ &  & \frac{Q^{2}\, G_{N}\left(t\right)^{2}}{t}\left(m_{N}^{2}\left(\frac{1}{\left(k_{2}\cdot k_{5}\right)^{2}}+\frac{1}{\left(k_{4}\cdot k_{5}\right)^{2}}\right)-\frac{2m_{N}^{2}-t^{\prime}}{\left(k_{2}\cdot k_{5}\right)\left(k_{4}\cdot k_{5}\right)}\right)+\label{br1.15gz}\\
\nonumber \\ &  & \frac{Q\, G_{N}\left(t\right)G_{N}\left(t^{\prime}\right)\left(t+t^{\prime}\right)}{t\, t^{\prime}}\left(\frac{m_{e}^{2}+m_{N}^{2}-u}{\left(k_{1}\cdot k_{5}\right)\left(k_{4}\cdot k_{5}\right)}-\frac{s-m_{e}^{2}-m_{N}^{2}}{\left(k_{1}\cdot k_{5}\right)\left(k_{2}\cdot k_{5}\right)}-\right.\nonumber \\
\nonumber \\ &  & \qquad\left.\left.\frac{s^{\prime}-m_{e}^{2}-m_{N}^{2}}{\left(k_{3}\cdot k_{5}\right)\left(k_{4}\cdot k_{5}\right)}+\frac{m_{e}^{2}+m_{N}^{2}-u^{\prime}}{\left(k_{2}\cdot k_{5}\right)\left(k_{3}\cdot k_{5}\right)}\right)\right).\nonumber \end{eqnarray}
 The scalar products $\left(k_{i}\cdot k_{5}\right)$ are Lorentz
invariants, and can be replaced with the Mandelstam variables as \begin{eqnarray}
\left(k_{1}\cdot k_{5}\right) & = & -m_{e}^{2}-m_{N}^{2}+\frac{s+t+u}{2},\nonumber \\
\nonumber \\\left(k_{2}\cdot k_{5}\right) & = & -m_{e}^{2}-m_{N}^{2}+\frac{s+t^{\prime}+u^{\prime}}{2},\nonumber \\
\label{br1.16}\\\left(k_{3}\cdot k_{5}\right) & = & m_{e}^{2}+m_{N}^{2}-\frac{s^{\prime}+t+u^{\prime}}{2},\nonumber \\
\nonumber \\\left(k_{4}\cdot k_{5}\right) & = & m_{e}^{2}+m_{N}^{2}-\frac{s^{\prime}+t^{\prime}+u}{2}.\nonumber \end{eqnarray}
 As for $\left|M_{a,Z}^{\prime}\right|_{L,R}^{2}$ and $Re(M_{a,\gamma}^{\prime}M_{a,Z}^{\prime})_{L,R}$,
we have detailed expressions given in the appendix of this article.
The helicity matrix elements were computed with the help of $FormCalc$
\cite{Hah97}.

Now we are ready to proceed to the next sections, where we shall give
the details on the parametrization of the emitted photon's phase space,
and numerical details on the calculations of the PV HPB contribution.

\section{HPB Differential Cross Section}

\begin{figure}[h]
\includegraphics[scale=0.8]{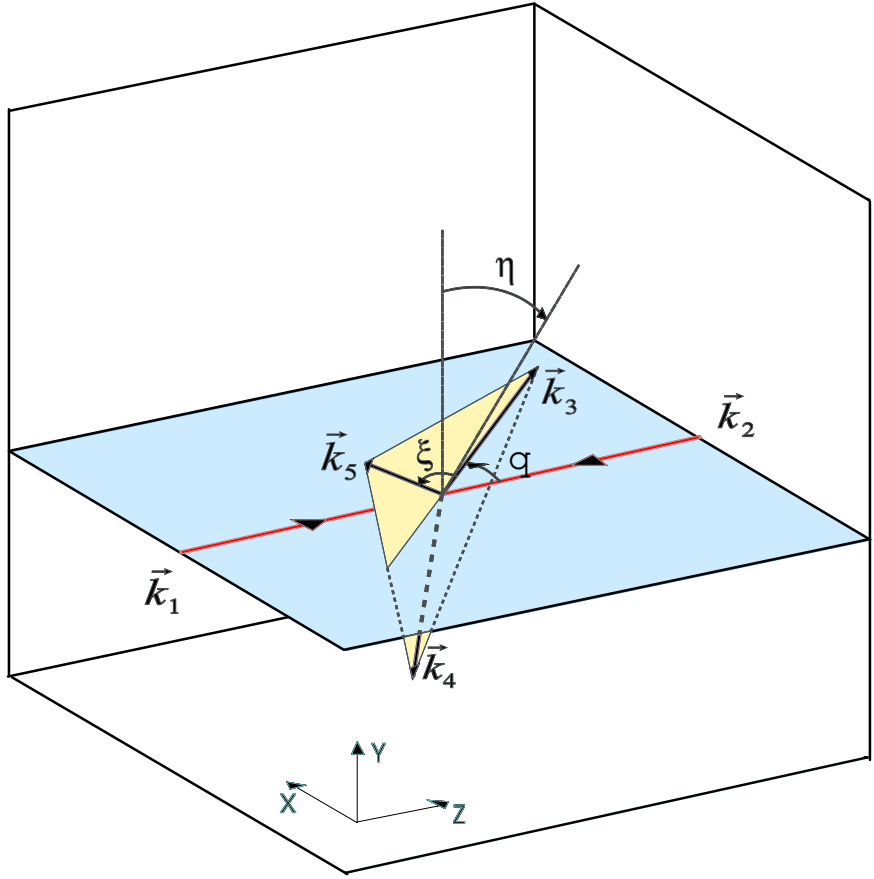} 

\caption{Phase space for the emitted hard photon.\label{fbr1.2} }
\end{figure}

Parametrization of the phase space for $(2\rightarrow3)$ process
has been chosen according to Fig. (\ref{fbr1.2}).

Here, the angle $\theta$ is a scattering angle and $\xi$ corresponds
to the angle between emitted photon and scattered electron. The momenta
are represented as \begin{eqnarray}
k_{1} & = & \left\{ E_{1},0,0,p_{in}\right\} ,\nonumber \\
k_{2} & = & \left\{ E_{2},0,0,-p_{in}\right\} ,\nonumber \\
k_{3} & = & \left\{ k_{3}^{0},\left|\overrightarrow{k_{3}}\right|\overrightarrow{e_{3}}\right\} ,\label{br1.35}\\
k_{4} & = & \left\{ k_{4}^{0},\overrightarrow{k_{4}}\right\} ,\nonumber \\
k_{5} & = & \left\{ k_{5}^{0},\left|\overrightarrow{k_{5}}\right|\overrightarrow{e_{5}}\right\} ,\nonumber \end{eqnarray}
 where unit vectors are \begin{eqnarray}
\overrightarrow{e_{3}} & = & \left(\begin{tabular}{c}
 sin$\left(\theta\right)$ \\
0 \\
cos$\left(\theta\right)$\end{tabular}\right),\nonumber \\
\label{br1.36}\\\overrightarrow{e_{5}} & = & \left(\begin{tabular}{c}
 cos$\left(\theta\right)\cos\left(\eta\right)\sin$$\left(\xi\right)+$ sin$\left(\theta\right)$ $\cos\left(\xi\right)$ \\
sin$\left(\eta\right)\sin\left(\xi\right)$ \\
$\cos\left(\theta\right)$ $\cos\left(\xi\right)$ $-\sin$$\left(\theta\right)$ $\cos\left(\eta\right)$ $\sin\left(\xi\right)$\end{tabular}\right).\nonumber \end{eqnarray}
 For on-shell particles, the incident momentum $p_{in}$ can be found
as \begin{equation}
p_{in}=\sqrt{E_{1}-m_{e}^{2}},\label{br1.36a}\end{equation}
 with \begin{equation}
E_{1}=\frac{E_{cms}+m_{e}^{2}-m_{N}^{2}}{2E_{cms}}.\label{br1.36aa}\end{equation}
 The center-of-mass energy $E_{cms}$ can be determined as follows:
\begin{equation}
E_{cms}=\sqrt{m_{e}^{2}+m_{N}^{2}+2E_{lab}\, m_{N}}.\label{br1.36b}\end{equation}

Momentum $k_{4}$ is determined by the four-momentum conservation
law in the cms frame: \begin{equation}
\sqrt{s}=k_{1}^{0}+k_{2}^{0}=k_{3}^{0}+k_{4}^{0}+k_{5}^{0},\label{br1.37a}\end{equation}
 and \begin{equation}
\overrightarrow{k_{3}}+\overrightarrow{k_{4}}+\overrightarrow{k_{5}}=0.\label{br1.38b}\end{equation}

The HPB differential cross section reads as follows \begin{equation}
d\sigma=\frac{\left|M_{tot}^{2\rightarrow3}\right|^{2}}{\Phi}d\Gamma^{3},\label{br1.38}\end{equation}
 where $\Phi$ is a flux factor and given by \[
\Phi=4p_{in}\sqrt{s}.\]
 The $(2\rightarrow3)$ process phase-space element $d\Gamma^{3}$
is \begin{equation}
d\Gamma^{3}=\frac{d^{3}k_{3}}{\left(2\pi\right)^{3}2k_{3}^{0}}\,\frac{d^{3}k_{4}}{\left(2\pi\right)^{3}2k_{4}^{0}}\,\frac{d^{3}k_{5}}{\left(2\pi\right)^{3}2k_{5}^{0}}\left(2\pi\right)^{4}\delta^{(4)}\left(k_{1}+k_{2}-k_{3}-k_{4}-k_{5}\right).\label{br1.39}\end{equation}
 Using \[
\frac{d^{3}k_{i}}{2k_{i}^{0}}=d^{4}k_{i}\delta\left(k_{i}^{2}-m_{i}^{2}\right)=\frac{\left|\overrightarrow{k_{i}}\right|}{2}dk_{i}^{0}d\Omega_{i},\]
 and the fact that the photon is a massless boson, i.e $\left|\overrightarrow{k_{5}}\right|=k_{5}^{0}$,
we can write \begin{equation}
d\Gamma^{3}=\frac{\left|\overrightarrow{k_{3}}\right|k_{5}^{0}}{4\left(2\pi\right)^{5}}dk_{3}^{0}d\Omega_{3}dk_{5}^{0}d\Omega_{5}\delta\left(k_{4}^{2}-m_{N}^{2}\right)d^{4}k_{4}\delta^{(4)}\left(k_{1}+k_{2}-k_{3}-k_{4}-k_{5}\right).\label{br1.40}\end{equation}
 Using the delta function $\delta^{(4)}\left(k_{1}+k_{2}-k_{3}-k_{4}-k_{5}\right)$
to eliminate the integration over momentum $k_{4}$, we arrive at
\begin{equation}
d\Gamma^{3}=\frac{\left|\overrightarrow{k_{3}}\right|\left|\overrightarrow{k_{5}}\right|}{4\left(2\pi\right)^{5}}dk_{3}^{0}d\Omega_{3}dk_{5}^{0}d\Omega_{5}\delta\left(k_{4}^{2}-m_{N}^{2}\right),\label{br1.40b}\end{equation}
 with $d\Omega_{3}=d\cos\theta\, d\phi$ and $d\Omega_{5}=d\cos\xi\, d\eta$.
The remaining delta function $\delta\left(k_{4}^{2}-m_{N}^{2}\right)$
will be used to eliminate integration over the scattered electron
energy $k_{3}^{0}.$

We need to do some modifications first: \begin{eqnarray}
k_{4}^{2}-m_{N}^{2} & = & \left(k_{4}^{0}\right)^{2}-\left|\overrightarrow{k_{4}}\right|^{2}-m_{N}^{2}\nonumber \\
 & = & \left(\sqrt{s}-k_{3}^{0}-k_{5}^{0}\right)^{2}-\left|\overrightarrow{k_{3}}\right|^{2}-\left|\overrightarrow{k_{5}}\right|^{2}-2\left|\overrightarrow{k_{3}}\right|\left|\overrightarrow{k_{5}}\right|\cos\left(\xi\right)-m_{N}^{2}.\label{br1.41a}\end{eqnarray}
 Now, using \begin{equation}
\left(k_{i}^{0}\right)^{2}=\left|\overrightarrow{k_{i}}\right|^{2}+m_{i}^{2},\label{br1.41b}\end{equation}
 we arrive at \begin{equation}
k_{4}^{2}-m_{N}^{2}=s-2\sqrt{s}k_{5}^{0}+m_{e}^{2}-m_{N}^{2}-2k_{3}^{0}\left(\sqrt{s}-k_{5}^{0}\right)-2\left|\overrightarrow{k_{3}}\right|k_{5}^{0}\cos\left(\xi\right).\label{br1.41c}\end{equation}
 The electron mass can be considered as a small parameter with respect
to $k_{3}^{0}$. In this case, we replace $\left|\overrightarrow{k_{3}}\right|$
by $k_{3}^{0},$ so that \begin{equation}
\left|\overrightarrow{k_{3}}\right|\simeq k_{3}^{0}-\frac{m_{e}^{2}}{2k_{3}^{0}}.\label{br1.42}\end{equation}
 Substitution of Eq.(\ref{br1.42}) into Eq.(\ref{br1.41c}) leads
to the following: \begin{equation}
k_{4}^{2}-m_{N}^{2}=\left(s-2\sqrt{s}k_{5}^{0}+m_{e}^{2}-m_{N}^{2}\right)-2k_{3}^{0}\left(\sqrt{s}-k_{5}^{0}+k_{5}^{0}\cos\left(\xi\right)\right)+\frac{m_{e}^{2}k_{5}^{0}\cos\left(\xi\right)}{k_{3}^{0}}.\label{br1.43}\end{equation}
 The property of the delta function $\delta\left[g\left(k_{3}^{0}\right)\right]=\sum_{i}\frac{\delta\left(k_{3}^{0}-r_{i}\right)}{\left|g^{\prime}\left(r_{i}\right)\right|}$
($r_{i}$ is $i$-th root of the equation $k_{4}^{2}-m_{N}^{2}=0$,
solved with respect to $k_{3}^{0}$) makes it possible to replace
$\delta\left(k_{4}^{2}-m_{N}^{2}\right)$ by \begin{equation}
\delta\left(k_{4}^{2}-m_{N}^{2}\right)=\frac{1}{2\left(\sqrt{s}+k_{5}^{0}\left(\cos\left(\xi\right)-1\right)\right)+\frac{m_{e}^{2}k_{5}^{0}\cos\left(\xi\right)}{r^{2}}}\delta\left(k_{3}^{0}-r\right),\label{br1.44}\end{equation}
 where \[
r=\frac{\left(s-2\sqrt{s}k_{5}^{0}+m_{e}^{2}-m_{N}^{2}\right)}{2\left(\sqrt{s}+k_{5}^{0}\left(\cos\left(\xi\right)-1\right)\right)}+\frac{m_{e}^{2}k_{5}^{0}\cos\left(\xi\right)}{\left(s-2\sqrt{s}k_{5}^{0}+m_{e}^{2}-m_{N}^{2}\right)}.\]
 The delta function $\delta\left(k_{3}^{0}-r\right)$ will eliminate
integration over $k_{3}^{0}$ leaving $k_{3}^{0}=r$. Integration
over the emitted photon's phase space $dk_{5}^{0}d\Omega_{5}$ can
be performed numerically using the cuts on the photon's energy $k_{5}^{0}:$
\begin{eqnarray}
\left(k_{5}^{0}\right)_{\min} & = & \Delta E,\nonumber \\
\label{br1.45}\\\left(k_{5}^{0}\right)_{\max} & = & \frac{\sqrt{s}}{2}-\frac{\left(m_{e}+m_{N}\right)^{2}}{2\sqrt{s}}.\nonumber \end{eqnarray}
 Finally, Eq. (\ref{br1.38}) becomes \begin{equation}
d\sigma_{R,L}^{HPB}=\frac{1}{4\left(2\pi\right)^{5}}\frac{1}{\Phi}\left(\intop\intop\intop\frac{\left|\overrightarrow{k_{3}}\right|\, k_{5}^{0}\,\left|M_{tot}^{2\rightarrow3}\right|_{R,L}^{2}dk_{5}^{0}d\Omega_{5}}{2\left(\sqrt{s}+k_{5}^{0}\left(\cos\left(\xi\right)-1\right)\right)+\frac{m_{e}^{2}k_{5}^{0}\cos\left(\xi\right)}{\left(k_{3}^{0}\right)^{2}}}\right)d\Omega_{3},\label{br1.46}\end{equation}
 leaving the final differential cross section differential with respect
to the scattered electron solid angle $d\Omega_{3}.$

\subsection{Numerical Test}

Now we can introduce details of the treatment of infrared divergences
in the parity-violating formfactors $C_{1,2}^{LO+NLO}$. Since the
PV amplitude derived from the Hamiltonian Eq.(\ref{eq:PVhamiltonian})
can be used in the calculations of the asymmetry Eq.(\ref{eq:pv1}),
the differential cross section for the neutral current reaction can
be computed as: \begin{equation}
d\sigma_{R,L}^{Z}=\frac{1}{4\left(2\pi\right)^{2}}\frac{1}{\Phi}\frac{\left|\overrightarrow{k_{3}}\right|}{\sqrt{s}}\left|M_{LO}^{Z,2\rightarrow2}+M_{NLO}^{Z,2\rightarrow2}\right|_{R,L}^{2}d\Omega_{3}.\label{br1.48}\end{equation}

The contribution of the soft- and hard-photon bremsstrahlung modifies
differential cross sections in the Eq.(\ref{eq:pv1}) according to
the following: \begin{equation}
d\widetilde{\sigma}_{R,L}^{Z}=d\sigma_{R,L}^{Z}+\left(d\sigma_{LO}^{Z,2\rightarrow2}\right)_{R,L}\cdot\delta_{soft}+d\sigma_{R,L}^{Z,HPB}.\label{br1.49}\end{equation}
 In order to combine HPB differential cross section with the soft-photon
emission contribution factor $\delta_{soft}$, and all this with parity-violating
formfactors $C_{1,2}^{LO+NLO}$, we propose the following parametrization
for the HPB differential cross section: \begin{equation}
d\widetilde{\sigma}_{R,L}^{Z,HPB}=\left(d\sigma_{LO}^{Z,2\rightarrow2}\right)_{R,L}\cdot\frac{d\sigma_{R}^{Z,HPB}-d\sigma_{L}^{Z,HPB}}{\left(d\sigma_{LO}^{Z,2\rightarrow2}\right)_{R}-\left(d\sigma_{LO}^{Z,2\rightarrow2}\right)_{L}}=\left(d\sigma_{LO}^{Z,2\rightarrow2}\right)_{R,L}\cdot\widetilde{\delta}_{HPB}.\label{br1.50}\end{equation}
 As can be easily seen, the substitution of Eq.(\ref{br1.50}) into
the expression for asymmetry Eq.(\ref{eq:pv1}) will leave terms related
to the neutral current reaction HPB in the usual form $\left(d\sigma_{R}^{Z,HPB}-d\sigma_{L}^{Z,HPB}\right)$.
It is worth noting that the term $\left(d\sigma_{R}^{tot}+d\sigma_{L}^{tot}\right)$
has a dominant contribution from the parity-conserving part of the
differential cross section. Because of that, denominator of Eq.(\ref{eq:pv1})
is left without parity-violating soft- and hard-photon bremsstrahlung
terms. Combining the soft term of Eq.(\ref{br1.6}) and the HPB term
of Eq.(\ref{br1.50}) with PV formfactors, we can write \begin{equation}
\widetilde{C}_{1,2}^{NLO}=C_{1,2}^{NLO}+\frac{C_{1,2}^{LO}}{2}\left(\delta_{soft}+\widetilde{\delta}_{HPB}\right).\label{br1.51}\end{equation}

For the case of $\left\{ e-N\right\} $ scattering, we will show numerical
contribution from the SPB and HPB terms, taking into account only
the IR finite part of the soft-photon bremsstrahlung only. We can
do so because IR divergences are canceled when $\left\{ e-N\right\} $
PV formfactors $C_{1,2}^{NLO}$ are combined with the second part
of the Eq.(\ref{br1.51}). Moreover we will treat formfactor $G_{N}\left(q\right)$
using monopole approximation ($n=1$) in our numerical tests.

Let us start with demonstration that, indeed, we do not have a $\Delta E$
dependence in the term $\frac{1}{2}\left(\delta_{soft}+\widetilde{\delta}_{HPB}\right)$
for a kinematic point relevant to the $Q_{weak}$ experiment. We take
$E_{lab}=1.165$ GeV and $Q^{2}=0.03\,$ GeV$^{2}$. During the numerical
integration, we have used the adaptive Genz-Malik algorithm which
is implemented in the $\emph{Mathematica}$ program \cite{Math}.
For electron-proton scattering, the term $\frac{1}{2}\left(\delta_{soft}+\widetilde{\delta}_{HPB}\right)$
for different values of $\Delta E$ is shown in the Table (\ref{t1.1}).%
\begin{table}
\begin{centering}
\begin{tabular}{|cc|}
\hline 
\multicolumn{1}{|c|}{$\Delta E$ $(\sqrt{s})$}&
$\frac{1}{2}\left(\delta_{soft}+\widetilde{\delta}_{HPB}\right)$ \tabularnewline
\hline
\hline 
$10^{-3}$ &
$-0.16966$ \tabularnewline
$10^{-4}$ &
$-0.16980$\tabularnewline
$10^{-5}$ &
$-0.16982$ \tabularnewline
$10^{-6}$ &
$-0.16984$ \tabularnewline
$10^{-7}$ &
$-0.16982$ \tabularnewline
\hline
\end{tabular}
\par\end{centering}

\caption{Dependence on the photon detector acceptance (electron-nucleon scattering
case $E_{lab}=1.165$ GeV, $Q^{2}=0.03\,$ GeV$^{2}$)\label{t1.1}}
\end{table}
We see that the variation of $\frac{1}{2}\left(\delta_{soft}+\widetilde{\delta}_{HPB}\right)$
is of order of $\sim0.1\%$ which is coming from the statistical error
of integration. The same can be done in the analysis of the $\Delta E$
dependence for the PV asymmetry due to soft and hard photon bremsstrahlung
(see Eqs.(\ref{eq:asymbr}), (\ref{br1.6}) and (\ref{br1.46})).%
\begin{table}
\begin{centering}
\begin{tabular}{|cc|}
\hline 
\multicolumn{1}{|c|}{$\Delta E$ $(\sqrt{s})$}&
$A_{B}(10^{-8}/d\Gamma^{2})$ \tabularnewline
\hline
\hline 
$10^{-3}$ &
$1.00019$ \tabularnewline
$10^{-4}$ &
$1.00061$\tabularnewline
$10^{-5}$ &
$1.00068$ \tabularnewline
$10^{-6}$ &
$1.00068$ \tabularnewline
$10^{-7}$ &
$1.00067$ \tabularnewline
\hline
\end{tabular}
\par\end{centering}

\caption{Dependence of the bremsstrahlung asymmetry given in the units of
$1/d\Gamma^{2}=\frac{4(2\pi)^{2}\sqrt{s}}{\left|\protect\overrightarrow{k_{3}}\right|}$
on the photon detector acceptance $\Delta E$ (electron-nucleon scattering
case $E_{lab}=3.0$ GeV, $Q^{2}=0.3\,$ GeV$^{2}$)\label{t1.2}}
\end{table}
Data for table (\ref{t1.2}) have been computed using the same integration
technique, and it is clear that variations of the asymmetry are $<0.05\,\%$.
In the test of independence from the $\Delta E$ parameter we took
one of the kinematic points of the G0 experiment, with $Q^{2}=0.3\,$
GeV$^{2}$. For the complete analysis of $e-p$ PV scattering asymmetries
it is required to include all the LO and NLO contributions, which
will be left to a future publication using the treatment of IR divergences
described in this article.

\section{Conclusion}

The calculation routines, tested for electron-proton scattering and
presented in our previous work \cite{BAB2002}, are valid for any
electroweak processes involving particles of the Standard Model. The
HPB contribution computed in the current work can be applied for virtually
any scattering process. Again, when computing hard-photon bremsstrahlung
terms for electron-proton scattering, we use effective $Z-p$ and
$\gamma-p$ couplings with monopole type form factors.

The enormous size of the complete analytical expressions involved
makes it impossible to present them in this paper. The complete analytical
expression in the $Mathematica$ file is available from authors upon
request.

We observed that for the energy range employed by the PV experiments
the HPB differential cross section is dominated by part of HPB amplitude
without the photon momentum in the numerator. It is still necessary
to keep in mind that the process is $(2\rightarrow3)$ when the cross
section is calculated. Terms proportional to $O\left(k_{5}\right)$
tend to be important for higher energies. This simplifies calculations
of the HPB cross section for the considered experiments significantly,
as it simplifies the numerator algebra.

We split the amplitude in two parts, with one part being the amplitude
without the momentum of the emitted photon in the numerator. This
step is important, because, according to the numerical analysis performed,
this term has a strong dominant structure similar to the soft-photon
emission factor. Another interesting result of this work is that all
of the effects, including soft- and hard-photon bremsstrahlung terms,
can be now accounted for on the level of PV formfactors.

The proper account of the soft- and hard-photon bremsstrahlung effects
has allowed us to achieve final results that are free from a logarithmic
dependence on the detector photon acceptance parameter.

\acknowledgments

The authors thank Malcolm Butler of Saint Mary's University for useful
comments. This work has been supported by NSERC (Canada). S. Barkanova
would also like to express her gratitude to Acadia University for
the generous start-up grant financing the part of this project.

\section{Appendix}

\begin{flushleft}
Here we give the detailed $\left|M_{a,Z}^{\prime}\right|_{L,R}^{2}$
expressions used for the calculations of $\left|M_{a,Z}^{2\rightarrow3}\right|_{L,R}^{2}$
in Eq.(\ref{br1.15}) for left-handed incident electrons: \begin{eqnarray}
 &  & \left|M_{a,Z}^{\prime}\right|_{L}^{2}=\frac{4\alpha^{3}\pi^{3}\left(1-2s_{W}^{2}\right)^{2}}{c_{W}^{2}\, s_{W}^{2}}\left[\frac{\left(f_{2}(0)\right)^{2}}{4m_{N}^{2}}(16m_{N}^{6}-8m_{N}^{4}(s+s^{\prime}+u+u^{\prime})+4m_{N}^{2}(s+u)(s^{\prime}+u^{\prime})+\right.\nonumber \\
\nonumber \\ &  & t^{\prime}(us^{\prime}-ss^{\prime}+tt^{\prime}+su^{\prime}-uu^{\prime}))+g_{R}^{Z-N}f_{2}(0)(4m_{N}^{4}-4m_{N}^{2}(u+u^{\prime})-ss^{\prime}+tt^{\prime}+s^{\prime}u+su^{\prime}+\nonumber \\
\label{br1.17}\\ &  & 3uu^{\prime})+2\left(g_{L}^{Z-N}\right)^{2}(m_{N}^{2}-s)(m_{N}^{2}-s^{\prime})+2\left(g_{R}^{Z-N}\right)^{2}(m_{N}^{2}-u)(m_{N}^{2}-u^{\prime})+\nonumber \\
\nonumber \\ &  & \left.g_{L}^{Z-N}(4g_{R}^{Z-N}m_{N}^{2}\,\, t+f_{2}(0)(4m_{N}^{4}-4m_{N}^{2}(s+s^{\prime})+3ss^{\prime}+tt^{\prime}+s^{\prime}u+su^{\prime}-uu^{\prime}))\right],\nonumber \end{eqnarray}
 and for the right-handed incident electrons: \begin{eqnarray}
 &  & \left|M_{a,Z}^{\prime}\right|_{R}^{2}=\frac{16\alpha^{3}\pi^{3}s_{W}^{2}}{c_{W}^{2}}\left[\frac{\left(f_{2}(0)\right)^{2}}{4m_{N}^{2}}(16m_{N}^{6}-8m_{N}^{4}(s+s^{\prime}+u+u^{\prime})+4m_{N}^{2}(s+u)(s^{\prime}+u^{\prime})+\right.\nonumber \\
\nonumber \\ &  & t^{\prime}(us^{\prime}-ss^{\prime}+tt^{\prime}+su^{\prime}-uu^{\prime}))+g_{R}^{Z-N}f_{2}(0)(4m_{N}^{4}-4m_{N}^{2}(s+s^{\prime})-uu^{\prime}+tt^{\prime}+s^{\prime}u+su^{\prime}+\nonumber \\
\label{br1.18}\\ &  & 3ss^{\prime})+2\left(g_{R}^{Z-N}\right)^{2}(m_{N}^{2}-s)(m_{N}^{2}-s^{\prime})+2\left(g_{L}^{Z-N}\right)^{2}(m_{N}^{2}-u)(m_{N}^{2}-u^{\prime})+\nonumber \\
\nonumber \\ &  & \left.g_{L}^{Z-N}(4g_{R}^{Z-N}m_{N}^{2}\,\, t+f_{2}(0)(4m_{N}^{4}-4m_{N}^{2}(u+u^{\prime})+3uu^{\prime}+tt^{\prime}+s^{\prime}u+su^{\prime}-ss^{\prime}))\right].\nonumber \end{eqnarray}
 First part $Re(M_{a,\gamma}^{\prime}M_{a,Z}^{\prime})_{L,R}$ of
the interference term Eq.(\ref{br1.15gz}) has the following structure
for the left-handed incident electrons
\par\end{flushleft}

\begin{flushleft}
\hfill{} \begin{eqnarray}
 &  & Re(M_{a,\gamma}^{\prime}M_{a,Z}^{\prime})_{L}=\frac{8\alpha^{3}\pi^{3}(1-2s_{W}^{2})}{c_{W}s_{W}}\left[g_{L}^{Z-N}(2(1+F_{2}(0))m_{N}^{4}+\right.\nonumber \\
\nonumber \\ &  & 2(t-(1+F_{2}(0))(s+s^{\prime}))m_{N}^{2}+s(\frac{3}{2}s^{\prime}+2s^{\prime}+\frac{1}{2}u^{\prime})+\frac{1}{2}F_{2}(0)(tt^{\prime}+s^{\prime}u-uu^{\prime}))+\nonumber \\
\label{eq:intL}\\ &  & g_{R}^{Z-N}(2(1+F_{2}(0))m_{N}^{4}+2(t-(1+F_{2}(0))(u+u^{\prime}))m_{N}^{2}+2uu^{\prime}+\frac{1}{2}F_{2}(0)(us^{\prime}-ss^{\prime}+\nonumber \\
\nonumber \\ &  & tt^{\prime}+su^{\prime}+3uu^{\prime}))+f_{2}(0)(4m_{N}^{4}-2(s+s^{\prime}+u+u^{\prime})m_{N}^{2}+tt^{\prime}+(s+u)(s^{\prime}+u^{\prime}))+\nonumber \\
\nonumber \\ &  & \left.\frac{f_{2}(0)F_{2}(0)}{4m_{N}^{2}}(16m_{N}^{6}-8(s+s^{\prime}+u+u^{\prime})m_{N}^{4}+4(s+u)(s^{\prime}+u^{\prime})m_{N}^{2}+t^{\prime}(tt^{\prime}-(s-u)(s^{\prime}-u^{\prime})))\right]\nonumber \end{eqnarray}
 and for right-handed electrons we have \begin{eqnarray}
 &  & Re(M_{a,\gamma}^{\prime}M_{a,Z}^{\prime})_{R}=-\frac{16\alpha^{3}\pi^{3}s_{W}}{c_{W}}\left[g_{R}^{Z-N}(2(1+F_{2}(0))m_{N}^{4}+\right.\nonumber \\
\nonumber \\ &  & 2(t-(1+F_{2}(0))(s+s^{\prime}))m_{N}^{2}+s(\frac{3}{2}s^{\prime}+2s^{\prime}+\frac{1}{2}u^{\prime})+\frac{1}{2}F_{2}(0)(tt^{\prime}+s^{\prime}u-uu^{\prime}))+\nonumber \\
\label{eq:intR}\\ &  & g_{L}^{Z-N}(2(1+F_{2}(0))m_{N}^{4}+2(t-(1+F_{2}(0))(u+u^{\prime}))m_{N}^{2}+2uu^{\prime}+\frac{1}{2}F_{2}(0)(us^{\prime}-ss^{\prime}+\nonumber \\
\nonumber \\ &  & tt^{\prime}+su^{\prime}+3uu^{\prime}))+f_{2}(0)(4m_{N}^{4}-2(s+s^{\prime}+u+u^{\prime})m_{N}^{2}+tt^{\prime}+(s+u)(s^{\prime}+u^{\prime}))+\nonumber \\
\nonumber \\ &  & \left.\frac{f_{2}(0)F_{2}(0)}{4m_{N}^{2}}(16m_{N}^{6}-8(s+s^{\prime}+u+u^{\prime})m_{N}^{4}+4(s+u)(s^{\prime}+u^{\prime})m_{N}^{2}+t^{\prime}(tt^{\prime}-(s-u)(s^{\prime}-u^{\prime})))\right]\nonumber \end{eqnarray}

\par\end{flushleft}

For simplicity, we have introduced a set of coupling constants $g_{L,R}^{Z-N}$
defined as \begin{eqnarray}
g_{R,L}^{Z-N} & = & f_{1}(0)\pm g_{1}(0),\nonumber \\
\\\left(\Gamma_{Z-N}^{\mu}\right)^{\prime} & = & ie\left[g_{R}^{Z-N}\gamma^{\mu}\varpi_{+}+g_{L}^{Z-N}\gamma^{\mu}\varpi_{-}+\frac{i}{2m_{N}}\sigma^{\mu\rho}\left(k_{4}-k_{2}\right)_{\rho}\, f_{2}(0)\right],\nonumber \end{eqnarray}
 where $\varpi_{\pm}=\frac{1\pm\gamma_{5}}{2}$ are the chirality
projector operators.

\end{document}